\documentclass[epj]{webofc}
\usepackage[varg]{txfonts}   
\usepackage{lineno}
\DeclareGraphicsExtensions{.pdf,png,.jpg}

\newcommand{\etal}{et al., }

\begin{document}
%
%
\title{The Science Case for a Southern Wide Field of View Detector}
%
%

\author{Di Sciascio Giuseppe\inst{1}\fnsep\thanks{\email{giuseppe.disciascio@roma2.infn.it}}         
}

\institute{INFN - Roma Tor Vergata
          }

\abstract{%
EAS arrays are survey instruments able to monitor continuously all the overhead sky. 
Their sensitivity in the sub-TeV/TeV energy domain cannot compete with that of Cherenkov telescopes, but the wide field of view (about 2 sr) is ideal to complement directional detectors by performing unbiased sky surveys, by monitoring variable or flaring sources such as Active Galactic Nuclei (AGN) and to discover transients or explosive events (GRBs). 
Arrays are well suited to study extended sources, such as the Galactic diffuse emission, and to measure the spectra of Galactic sources at the highest energies (near or beyond 100 TeV).
An EAS array is able to detect at the same time events induced by photons and charged cosmic rays, thus studying the connection between these two messengers of the non-thermal Universe.
Therefore, these detectors are, by definition, multi-messenger instruments. 
All EAS arrays presently in operation or under installation are located in the Northern hemisphere.
The scientific potential of a next-generation survey instrument in the Southern Hemisphere will be presented and briefly discussed.
}
\maketitle
\section{Introduction}
\label{intro}

The riddle of the origin of Cosmic Rays (CR) is unsolved since more than one century.
Charged cosmic rays, gammas and neutrinos are strongly correlated in CR sources where hadronic accelerators are at work. 
Their integrated study is one of the most important and exciting fields in the \emph{'multi-messenger astronomy'}, the exploration of the Universe through combining information from different cosmic messengers: electromagnetic radiation, gravitational waves, neutrinos and cosmic rays.

The study of CRs is based on two complementary approaches:
\begin{enumerate}
\item[(1)] Measurement of energy spectrum, elemental composition and anisotropy in the CR arrival direction distribution, the three basic observables crucial for understanding origin, acceleration and propagation of the radiation.
\item[(2)] Search of their sources through the observation of neutral radiation (photons and neutrinos), which points back to the emitting sources not being affected by the magnetic fields.
\end{enumerate}
%

The identification of the galactic sources able to accelerate particles beyond PeV (=10$^{15}$ eV) energies, the so-called \emph{'PeVatrons'}, is certainly one of the main open problems of high energy astrophysics. In fact, even there is no doubt that galactic CR are accelerated in SuperNova Remnants (SNRs), the capability of SNRs to accelerate CRs up to the \emph{'knee'} of the all-particle energy spectrum ($\sim$3$\times$10$^{15}$ eV) and above is still under debate. The determination of the maximum energy at which protons are accelerated inside their sources (the \emph{'proton knee'}), as well as the measurement of the evolution of the heavier component across the knee, are the key for understanding acceleration mechanisms and the propagation processes in the Galaxy, and to investigate the transition from Galactic to extra-galactic CRs.

In a hadronic interaction the secondary photons have an energy a factor of 10 lower than the primary proton. 
Therefore, the quest for CR sources able to accelerate particles up to the knee energy region and above requires to survey the $\gamma$-ray sky above 100 TeV. 
In addition, the Inverse Compton scattering at these energies is strongly suppressed by the Klein-Nishina effect. Therefore, the observation of a $\gamma$-ray power law spectrum extending up to the 100 TeV range would be a strong indication of the hadronic nature of the emission.

So far no photons above 100 TeV have been observed from any source, and only a few sources have data above 30 TeV.
Their spectra is, however, only known with large uncertainties, being the sensitivity of the current instruments at the highest energies not enough to determine clearly the spectral shape. 

To open the 100 TeV range to observations a detector with a very large effective area, operating with high duty-cycle, is required.
The most sensitive experimental technique for the observation of $\gamma$-rays at these energies is the detection of EAS via large ground-based arrays.

EAS arrays are also extremely valid instruments in the sub-TeV/TeV energy domains.
Their sensitivity in this energy range cannot compete with that of the future generation Cherenkov telescopes, as CTA (that also have a better energy and angular resolution) \cite{cta-science}, but the wide FoV offers the opportunity to continuously monitor a large fraction of the sky, looking for unpredictable transient phenomena. 
In fact, most current Cherenkov telescopes can work only during clear moonless nights, with a total observation time of about 1000--1500 hours per year (depending on the location), with a typical duty cycle of $\sim$10--15$\%$, and a FoV of a few degrees radius. This implies that they can observe only one source at the time, and only in the season of the year when the source culminates during night time.

By contrast, an EAS detector every day can observe a large fraction of the celestial sphere (spanning 360$^{\circ}$ in right ascension and about 90$^{\circ}$ in declination, in a declination interval depending on the geographic location). 
Sources located in this portion of the sky are in the FoV of the detector, either always, or for several hours per day, depending on their celestial declination.
This situation is ideal to perform sky surveys, discover transients or explosive events, such as GRBs, and monitor variable or flaring sources such as AGN.

Another important topic that can be successfully addressed by a wide FoV detector is the extended $\gamma$-ray emission and in particular the study of the galactic diffuse gamma emission. This radiation
could trace the location of the CR sources and the distribution of interstellar gas.
In addition, the observation of a knee in the energy spectrum of the diffuse emission, at an energy of a few hundred TeV, corresponding to the knee of the CR all-particle spectrum, would provide a complementary way to investigate the origin of the knee \cite{vernetto-icrc17}. 
A new wide FoV experiment should also be able to map the diffuse emission in different regions of the Galactic Plane to observe a location dependence of the $\gamma$-ray spectral index \cite{argo-diffuse} and/or of the knee energy.

Wide FoV telescopes are also important for a multi-messenger study of the Gravitational Wave events due to their capability to survey simultaneously all the large sky regions identified by LIGO and VIRGO, looking for a possible correlated $\gamma$-ray emission.
As an example, the event GW170817 was localized within a 3 by 10 degree region, well beyond the typical FoV of a Cherenkov telescope and required multiple pointings to H.E.S.S. to cover the area \cite{hess-gw}.

\begin{figure}[ht]
\centering
\includegraphics[scale=0.30]{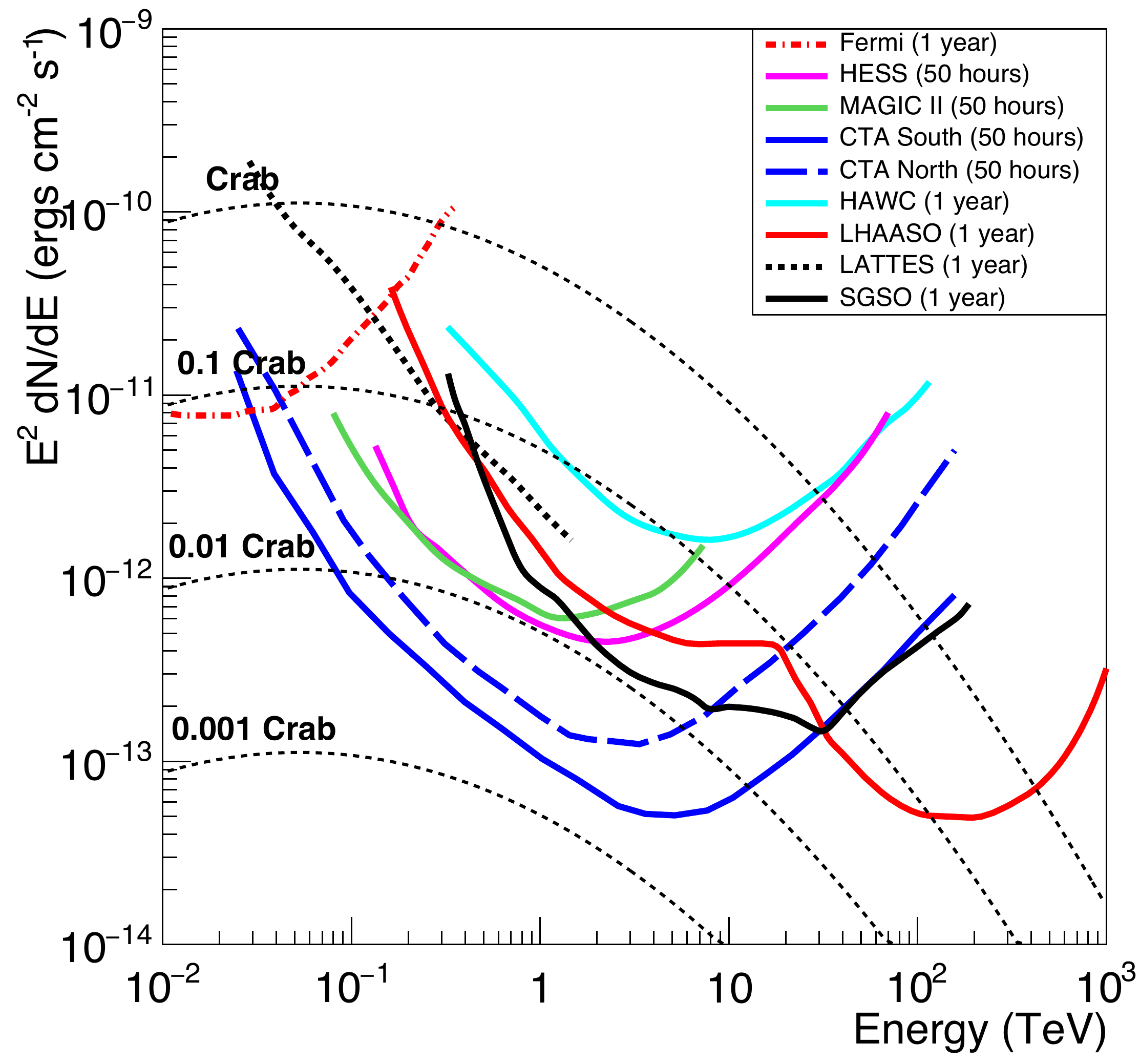}
\caption{Differential sensitivity (multiplied by E$^2$) to a Crab-like point gamma ray source of different experiments and projects. The best fit of the Crab Nebula data obtained by different detectors \cite{hhh15} is reported as a reference, extrapolated to 1 PeV}.
\label{fig:sensitivity}       
\end{figure}

The only wide FoV detector currently in data taking to study gamma-ray astronomy in the sub-TeV/multi-TeV energy domain is HAWC, located in central Mexico at an elevation of 4100 m a.s.l.  Two new wide FoV detectors, LHAASO and TAIGA-HiSCORE, are under installation. All these experiments are located in the Northern hemisphere. Differential sensitivity (multiplied by E$^2$) to a Crab-like point gamma ray source of different experiments and projects is shown in Fig. \ref{fig:sensitivity}.
%
%

The construction of a new survey instrument at sufficiently Southern latitude to continuously monitor the Galactic Center and the Inner Galaxy should be a high priority. 
Such an array could be also a 'finder' telescope for the future CTA-South experiment.

Science case and experimental solutions for a survey instrument in the South are under discussion in the framework of the Southern Gamma-ray Survey Observatory (SGSO) alliance \cite{sgso}. The sensitivity of a water Cherenkov based array covering an area of 221,000 m$^2$ with a fill-factor of 8\% is shown in Fig. \ref{fig:sensitivity}.

\subsection{The strong case for a wide FoV telescope in the Southern hemisphere}

The observation of the diffuse emission by AGILE \cite{agile-diffuse} and Fermi \cite{fermi-diffuse} shows that the distribution of CRs above the GeV energy range is smoothly distributed throughout the disc of the Galaxy but the intensity is higher in the Inner Galaxy and falls off towards the outer disc. 
Accordingly, the majority of $\gamma$-ray sources discovered in the last years above 100 GeV have been detected by the Cherenkov telescopes of the H.E.S.S. experiment located in Namibia.
A continuous monitoring of the Inner Galaxy is expected to allow the detection of a large number of new sources.

In addition to the reasons mentioned in the Introduction, the opportunity to observe the Inner Galaxy offers a lot of additional motivations to push the construction of a wide FoV in the Southern hemisphere.

As mentioned, H.E.S.S. recently claimed the possible detection of a PeVatrons in the Galactic Center, most likely related to a supermassive black hole \cite{hess-pev}, thus opening new perspectives for the observations of CR sources other than SNRs. In fact, the diffuse emission around J1745-290 (positionally compatible with SgrA$^*$) extends up to $\sim$50 TeV, thus suggesting acceleration of CR protons up to the PeV energy range.
Combined with the Fermi-LAT observations of a harder spectrum in the Galactic center region \cite{gaggero17} and the discovery of the so-called Fermi bubbles, we expect that a lot of high-energy particle accelerators are at work in this region that can be studied by a future wide FoV telescope.

Some authors found that the bulk of the Galactic Ridge emission can be naturally explained by the interaction of the diffuse steady-state Galactic CR sea with the gas present in the central molecular zone. Although they confirm the presence of a residual radial-dependent emission associated with a central source, the relevance of the large-scale diffuse component prevents claim solid evidence of a PeVatrons in the Galactic Center to explain the H.E.S.S. data \cite{gaggero17}.
A wide FoV detector may confirm this scenario observing the emission from larger region centered on the Galactic Center.

Another interesting opportunity for a wide FoV telescope located in the Southern hemisphere is to investigate the so-called \emph{'IceCube spectral anomaly'}. Recently it was suggested that different IceCube datasets are not consistent with the same power law spectrum of cosmic neutrinos, thus suggesting that they are observing a multicomponent spectrum. 
The passing-muon data (from the Northern sky) agree with isotropy and E$^{-2}$ distribution up to very high energies.
The HESE events (mostly from the Southern sky and shower-like) suggest a softer distribution at low energies. This North-South asymmetry is the IceCube spectral anomaly. It strengthens the likelihood that a Galactic neutrino component exists, mainly observable from the Southern hemisphere.  This Galactic component of neutrinos implies the existence of a Galactic component of very high energy gamma rays \cite{palladino16}.  
A measurement of the $\gamma$-ray flux from the Southern sky in the same energy domain of neutrinos (above 30 TeV) could clarify the origin of the emission. 

Recently, some authors emphasized the existence of an extended 'hot' region of the Southern gamma sky where the cumulative sources contribution dominates over the diffuse component \cite{pagliaroli17}. This region is located in the Inner Galactic Plane
and could be also an important source of HE neutrinos.
Interestingly, this region approximately coincides with the portion of the galactic plane from which a $\sim$2$\sigma$ excess of showers is observed in the HESE IceCube data sample.
If photons are not absorbed, we expect that the high energy neutrino sky is strongly correlated with the high energy gamma sky. A wide FoV detector will allow to study the gamma/neutrino correlation and will provide us a handle to perform a detailed multi-messenger study of the Galactic Plane.

With EAS array some aspects of fundamental physics can be also studied, in particular the existence of multi-TeV Dark Matter particles. The assumption of a single astrophysical power-law flux to explain the IceCube 6-year HESE events yields an anomalous large spectral index ($\gamma^{6yr}$=2.92). Adopting a spectral index in the range [2.0, 2.2], which is compatible with the one deduced by the analysis performed on the 6-year up-going muon neutrinos data, the latest IceCube data show an up to 2.6$\sigma$ excess in the number of events in the energy range 40--200 TeV. Such an excess can be interpreted as a decaying Dark Matter signal \cite{chianese17}. DM particles with masses in the 100 TeV energy range can be investigated only by large wide FoV detectors \cite{lhaaso-dm}.

Finally, a wide FoV detector is an ideal instrument to deepen the observations of the CR anisotropy carried out in the Southern hemisphere by IceCube/IceTop experiments \cite{iceaniso} as a function of the particle rigidity.

\section{Design consideration for a Southern Hemisphere Wide FoV detector}

A new Southern Hemisphere wide FoV detector, to be fully complementary to CTA-South and to carry out an integrated study of gamma and nuclei induced showers, needs 
\begin{enumerate}
\item an energy threshold of $\sim$100 GeV, to be a transient factory;
\item a sensitivity at few percent Crab flux level below the TeV, to have high exposure for flaring activity;
\item an angular resolution of $\sim$1$^{\circ}$ at the threshold, to reduce source confusion in the Inner Galaxy;
\item to survey the gamma sky at 100 TeV with a capability to discriminate the background at a level of 10$^{-5}$ to observe the knee in the energy spectrum of the gamma diffuse emission in different regions of the Galactic Plane;
\item to be able to discriminate different primary masses in the knee energy range to measure the proton knee and investigate the maximum energy of accelerated particles in CR sources and to observe the CR anisotropy as a function of the particle rigidity.
\end{enumerate}
%
%
Is this possible ?\\
The main parameters to push down the sensitivity to gamma-ray sources are \cite{discia-icrc17}: (1) the energy threshold; (2) the angular resolution; (3) the gamma/hadron relative trigger efficiency; (4) the effective area for photon detection; (5) the background rejection capability.

Two different experimental techniques have been applied in the last two decades in ground-based survey instruments: (1) water Cherenkov (Milagro); (2) Resistive Plate Chambers (ARGO-YBJ). 

The benefits in the use of RPCs in ARGO-YBJ were \cite{discia-rev}: (1) high efficiency detection of low energy showers (energy threshold $\sim$300 GeV at 4300 m asl) by means of the high density sampling of the central carpet ($\sim$92\% coverage); (2) unprecedented wide energy range investigated by means of the digital/charge read-outs ($\sim$300 GeV $\to$ 10 PeV, with a linearity of the read-out up to $\approx$10$^4$ p/m$^2$); (3) good angular resolution ($\sigma_{\theta}\approx 1.66^{\circ}$ at the threshold, without any lead layer on top of the RPCs) and unprecedented details in the core region by means of the high granularity of the different read-outs.

RPCs allowed to study also charged CR physics (energy spectrum, elemental composition and anisotropy) up to about 10 PeV.
By contrast, the capability of water Cherenkov facilities in extending the energy range to PeV and in selecting primary masses must be investigated.

In both experiments (Milagro and ARGO-YBJ) the limited capability to discriminate the background was mainly due to the small dimensions of the detectors. In fact, in the new experiments, like HAWC and LHAASO, the discrimination of the CR background is made studying shower characteristics far from the shower core (at distances R$>$ 40 m from the core, the dimension of the Milagro and ARGO-YBJ detectors).

The key to lower the energy threshold is to locate the detector at extreme altitude (about 5000 m asl for a threshold in the 100 GeV range). But the energy threshold, as well as the angular resolution, depends also on the coverage (the ratio between the detection area and the instrumented one), on the granularity of the read-out, on the particular type of detector and on the trigger logic.

The ARGO-YBJ experiment, combining the full coverage approach at high altitude with a high granularity of the read-out (about 15,000 strips 7$\times$62 cm$^2$ wide), sampled 100 GeV $\gamma$-induced showers with an efficiency of about 70\% at 4300 m a.s.l. . 
The median energy of the first multiplicity bin (20-40 fired pads) for photons with a Crab-like energy spectrum was 340 GeV \cite{argo-crab}.
The granularity of the read-out at cm level allowed to sample events with only 20 fired pads, out of 15,000, with a background-free topological-based trigger logic.
A water Cherenkov based facility will hardly be able to lower the energy threshold below $\sim$500 GeV. 

Therefore, a full coverage approach based on the RPC technology is one of the most interesting solution for a new survey instrument in the South. A hybrid detector with RPCs coupled to a HAWC/LHAASO - like water Cherenkov detector, will allow to lower the energy threshold at 100 GeV level, to measure the arrival direction with two independent techniques, to exploit the HAWC/LHAASO approach to reject the background, and to study also CR physics with nuclei induced events up to 10 PeV (the STACEX proposal).

\section{Conclusions} 

Open problems in cosmic ray physics push the construction of new generation EAS arrays to study, in the 10$^{11}$ -- 10$^{18}$ eV energy range, at the same time photon- and charged-induced events.
In the next decade CTA-North and LHAASO \cite{lhaaso1} are expected to be the most sensitive detectors to study gamma-ray astronomy in the Northern hemisphere from about 20 GeV up to PeV.

A survey instrument to monitor the Inner Galaxy and the Galactic Center with high sensitivity should be a high priority.
Extreme altitude ($\sim$5000 m asl), high coverage coupled to a high granularity of the read-out are the key to improve the angular resolution and the sensitivity to gamma-ray sources. A detector able to sample shower up to particle densities of $\approx$10$^4$ - 10$^5$ p/m$^2$ will allow to study primary particles up to $\sim$ 10 PeV.

The ARGO-YBJ Collaboration demonstrated that bakelite RPCs can be safely operated at extreme altitudes for many years. 
The benefits in the use of RPCs in ARGO-YBJ were: (1) high efficiency detection of low energy showers (energy threshold $\sim$300 GeV) by means of the dense sampling of the central carpet; (2) unprecedented wide energy range investigated by means of the digital/charge read-outs ($\sim$300 GeV $\to$ 10 PeV); (3) good angular resolution and unprecedented details in the core region by means of the high granularity of the read-outs.

The ARGO-like RPCs should be an important element of a future experiment in the South, possibly coupled to HAWC/LHAASO - like water Cherenkov detectors to exploit the added values of two experimental approaches. This is the aim of the STACEX proposal.

Science case and experimental solutions for a survey instrument in the South are under discussion in the framework of the Southern Gamma-ray Survey Observatory (SGSO) alliance. Science case and preliminary calculations of the sensitivity of a water Cherenkov based array, covering an area of 221,000 m$^2$ with a fill-factor of 8\%, are presented \cite{sgso} (see Fig. \ref{fig:sensitivity}).

\end{document}